\documentclass{article}

\usepackage{arxiv}

\usepackage[utf8]{inputenc} 
\usepackage[T1]{fontenc}    
\usepackage{hyperref}       
\usepackage{url}            
\usepackage{booktabs}       
\usepackage{amsfonts}       
\usepackage{nicefrac}       
\usepackage{microtype}      
\usepackage{lipsum}		
\usepackage{graphicx}
\usepackage[sort,numbers]{natbib}
\usepackage{doi}

\title{Tagged Documents Co-Clustering}
\author{
	\href{https://orcid.org/0000-0002-9072-1535}{\includegraphics[scale=0.06]{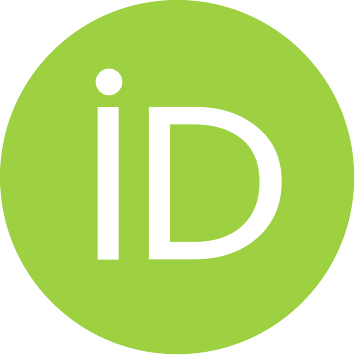}\hspace{1mm}Ga\"elle Candel} \\
	Wordline TSS Labs, Paris \\
	\texttt{firstname.lastname@worldline.com} \\
	\& \\
	D\'epartement d'informatique de l'ENS \\
	ENS, CNRS, PSL University, Paris \\
	\texttt{firstname.lastname@ens.fr}\\
	\And
  David Naccache \\
	D\'epartement d'informatique de l'ENS \\
	ENS, CNRS, PSL University, Paris \\
	\texttt{firstname.lastname@ens.fr}\\
}

\date{} 					

\renewcommand{\headeright}{}
\renewcommand{\undertitle}{}

\renewcommand{\shorttitle}{Tagged Documents Co-Clustering}

\hypersetup{
pdftitle={Tagged Documents Co-Clustering},
pdfsubject={cs.AI, cS.IR},
pdfauthor={Ga\"elle Candel, David Naccache},
pdfkeywords={Bipartite graphs, Hierarchical co-clustering, Power-law, Textual documents, Unsupervised learning},
}

\begin{document}
\maketitle

\begin{abstract}
	Tags are  short sequences of words allowing to describe textual and non-texual resources such as as music, image or book.
	Tags could be used by machine information retrieval systems to access quickly a document.
	These tags can be used to build recommender systems to suggest similar items to a user.
	However, the number of tags per document is limited, and often distributed according to a Zipf’s law.
	In this paper, we propose a methodology to cluster tags into conceptual groups.
	Data are preprocessed to remove power-law effects and enhance the context of low-frequency words. Then, a hierarchical agglomerative co-clustering  algorithm is proposed 	to group together the most related tags into clusters.
	The capabilities were evaluated on a sparse synthetic dataset and a real-world tag collection associated with scientific papers. The task being unsupervised, we propose some stopping criterion for selectecting an optimal partitioning.
\end{abstract}

\keywords{Bipartite graphs \and Hierarchical co-clustering \and Power-law \and Textual documents \and Unsupervised learning}

\section{Introduction}

Tags are words or short sequences associated with a resource or a document.
Depending on the context, the role of a tag differs.
They could be used to describe feeling, category, source, content, ownership and others \cite{surver_social_tagging}.

The tags associated with a document form a bag of words, where the order does not matter.
Tags can be weighted or ranked by relevance, helping the machine or user to know which are the most accurate or specific to the document.

Tags can be obtained with two different approaches:
they can be machine extracted, where an algorithm process the resources and extract some of their characteristics \cite{the_million_song_dataset,Tang:08KDD};
or they can be handcrafted by \textit{experts} or \textit{non-experts}.
In the expert case, the vocabulary is controlled, and the results would be comparable to machine extracted tags.
In contrast, \textit{folksonomy} corresponds to non-expert tags, leading to a large corpus with redundant or mispelt tags.

The corpus is often represented using the vector space model, where a document is represented by a sparse binary vector, where a $1$ encode the presence of a tag within the document.
The document's tags cannot be rank by importance because a document is described by a binary vector.
Tags can only be ranked within the corpus, and often shows power-law distribution \cite{power_tags}, making their analysis difficult because of the scarcity of frequent keywords and an abundance of unfrequent tags.
Additionally, their number per document is relatively smaller than for usual textual documents, as the goal of tags is to provide a synthetic view of the document.
This reduces the precision of an analysis when a tag is missing, as there is not enough tag redundancy to compensate for the absence of this tag.

A way to improve the tag vector representation is to rely on an external database \cite{clustering_short_sequence_wiki} like WordNet \cite{Wordnet} or Wikipedia \cite{wikipedia} to add additional related tags to enrich the initial tag vector description.
Another possibility is to rely on a probabilistic model \cite{DBSCAN_KL}, modeling keywords distribution based on available data.

Tags can be use for information retrieval and recommendation.
For a machine, it allows proposing related documents using tag co-occurrences.
For a human, tags can be used to create \textit{tag clouds} \cite{tag_cloud} to help a user to refine its query by suggesting related keywords.
A tag cloud displays the best co-occurring tags, adjusting the size, color, opacity of the tags to the context.
Nonetheless, the tag cloud utility is limited due to the amount of irrelevant unorganized information \cite{tag_cloud_signalers}, \cite{begelman2006automated}.
Rather than ranking tags against an initial query, another option is to cluster tags  \cite{tag_clusters}, grouping them by context.

Many algorithms, trying to cluster keywords alone, without clustering documents \cite{topic_detection_KL,keywords_clustering_autoclass,begelman2006automated} exist.
These approaches do not take advantage of the duality between samples and features.
For our particular setup where documents are succinctly described,
it seems more relevant to use an algorithm clustering  samples and features at the same time to improve the clustering quality.

Co-clustering approaches cluster both on rows and columns together.
Many approaches focus on the bipartite graph representation \cite{keywords_advertiser,co_clust_image_keywords,Dhillon01co-clusteringdocuments} of keywords and documents.
\cite{Dhillon01co-clusteringdocuments} proposed to use the spectral decomposition to cluster samples and features using their eigenvector representation.
The work \cite{Dhillon03InformationCoclustering} proposed using information theory methods, which given an initial partitioning alternates between row clusters refinement and column clusters refinement.

There are two main problems in clustering:
defining the target number of clusters, and defining the rules for cluster assignment.
Concerning the cluster count, we propose a hierarchical agglomerative algorithm that stores the merging operation history.
It free us from setting \textit{a priori} a specific number of clusters.
Nonetheless, we suggest a stopping criterion to select an optimal partitioning.
We follow the probabilistic and information theory approaches to cluster tags with limited information available.
We follow a co-clustering approach to take advantage of the synergies between documents and keywords clustering.

In this paper, we describe in the first part, a procedure to enhance keywords and documents context  without the use of an external database.
Next, the algorithm and its cost function are detailed.
The details about datasets and metrics used are presented in the experimental setup section, followed by the experimental results.
Then, the article ends with a discussion and a conclusion.

\section{Enlarging Document Context}

\subsection{Documents Keywords Matrix}

Be $X=\{x_i\}_{i=1:n}$ the set of $n$ documents and  $Y=\{y_j\}_{j=1:m}$ the set of $m$ keywords, which occur in the set of documents.
Using the vector space model, a document is represented under the form of a binary vector $\mathbf{x}_i = [x_{i, 1}, ..., x_{i, m}]$ where a $1$ encodes the keyword's presence while a $0$ its absence in document $i$.
The same representation applies to keywords,  represented as $\mathbf{y}_j = [y_{j, 1}, ..., y_{j, n}]$, where a $1$ encodes the occurrence of keyword $j$ in a document.

The collection of document vectors is often represented in a matrix form.
The documents-keywords matrix is $M \in \{0, 1\}^{n \times m}$, with $M_{i, j} = x_{i, j} = y_{j, i}$, where rows represent documents and columns the keywords.

\subsection{Similarity Matrix}

In practice, the documents-keywords matrix is very sparse as a document is assumed to be tagged with only a few relevant keywords.
For a keyword, the sparsity leads to very few keyword co-occurrence pairs with very low weight.

In our particular case, we assume that keywords are exact and relevant, accurately extracted by the algorithms or experts, and none due to mistakes.
This assumption simplifies the task of enhancing the keywords-documents matrix.
Rather than searching for incorrect pairs first for data cleaning,
all the keywords pairs are taken into account to improve the keywords-document matrix.

To improve the matrix, we take advantage of the co-occurrence matrices, one measuring the document's similarity, the other measuring the keyword's similarity.

A cosine-like similarity is used to compute the similarity between pairs of documents and keywords, $S^{(X)}$ and $S^{(Y)}$ respectively, taking values in $[0, 1]$.
Instead of running the cosine similarity on binary vector, the values are adjusted by the relative frequency.
For document $i$, each keyword is weighted by its frequency $c_k^{-1}$ where $c_k = \sum_{i} M_{i, k}$, to ensure that frequent keywords count less than infrequent ones.
The same normalization is performed on keyword vectors, normalized by the number of keywords within each document  $r_k = \sum_{j=1}^m M_{k,j}$.
The document normalization has almost no effect, as the number of tags per documents is relatively homogeneous, but affect a lot keyword normalization, as they follow a power-law distribution.

The document similarity is defined as:
\begin{equation}
 S^{(X)}_{i, j} = \frac{\sum_{k=1}^{m} \frac{M_{i, k} M_{j, k}}{c_k} }{\sqrt{\left(\sum_k^m \frac{M_{i, k}}{c_k}\right) \left( \sum_k^m \frac{M_{j, k}}{c_k}\right)}}
 \label{eq:cosine}
\end{equation}
The equation \ref{eq:cosine} is adapted to compute $S^{(Y)}$ by making the sum over the rows rather than the columns (i.e., $\sum_{k=1}^n \frac{M_{k, i}}{r_k}$), and normalized by the number of keywords describing the document $r_k$.
The normalization is equivalent to the inverse document frequency used in TF-IDF to score keywords by relevance, enabling to focus on singular keywords rather than frequent one.

\subsection{Transition Matrix}

The similarity matrix represents on a scale from $0$ to $1$ how well two elements are related.
The closeness is characterized by a value close to $1$ while unrelatedness by $0$.
Similar keywords can be either synonymous or occurring in the same context. In some way, the similarity matrix represents the co-occurrence laws.

To exploit these relationships, we suggest transforming the similarity matrices into the form of Markov transition probability matrices, where $T_{i, j} = Pr(j \rightarrow i)$ is the probability to move to state $j$ starting from $i$.
Using the product between  $T$ and $M$, we would obtain a vector with weights representing the probability of obtaining a given keyword given the initial tags.

A simple normalization of $S$ is not enough to obtain $T$.
The normalization of $S$ ensures that the mass-distributed by an item to its neighbors is preserved after transformation, ie $|T \mathbf{u}^{\dagger}| = |u|$.
However, it does not ensure that the mass is fairly attributed.
Highly frequent keywords are more receptive than infrequent ones, as $\sum_{j} Pr(j \rightarrow i)$ is larger for frequent items.

To rebalance the mass attribution, the matrix $T$ is obtained by bi-stochastization, leading to $T = T^{\dagger}$, ie $Pr(i \rightarrow j) = Pr(j \rightarrow i)$.
We use the Sinkhorn-Knopp algorithm \cite{SK_sto_binorm}, which alternates between searching the best column normalization vector $\mathbf{c}$ and the best row normalization vector $\mathbf{r}$, repeating until convergence $\mathbf{c} = (S \mathbf{r})^{-1}$ and $\mathbf{r} = (S^{\dagger} \mathbf{c})^{-1}$. The transition matrix is obtained by:
\begin{equation}
  T  = \mathcal{D}(\mathbf{r}) S \mathcal{D}(\mathbf{c})
  \label{eq:transition_matrix}
\end{equation}
where $\mathcal{D}(\mathbf{z})$ is the diagonal matrix with element $D_{i, i} = z_i$.

\subsection{Matrix Smoothing with Mass Preservation}

The two transition matrices are used to smooth the initial documents-keywords matrix using:
\begin{equation}
  M^* = T^{(X)} M T^{(Y)}
  \label{eq:matrix_smoothing}
\end{equation}
The detail of a term of  $M^*$ obtained following \ref{eq:matrix_smoothing} is:
\begin{equation}
    M^*_{i, j} = \sum_{k=1}^n \sum_{\ell=1}^m Pr(x_i | x_k) Pr(y_j | y_{\ell}) M_{k, \ell}
    \label{eq:detail_matrix_smoothing}
\end{equation}
The term $M^*_{i, j}$ of \ref{eq:detail_matrix_smoothing} corresponds to all possible transitions to a document $i$ and keywords $j$ from all possible document-keyword pairs $M_{k \ell}$.

The application of $T^{(Y)}$ preserves the mass on the rows while $T^{(X)}$ preserves it on columns. After application of both, only the global mass is preserved $\sum_{i,j} M^*_{i, j} = \sum_{i, j} M_{i, j}$.

This transformation redistributes the weights for documents and keywords without changing the global mass of the system.
The matrix $M^*$ will be used for the co-clustering instead of the binary matrix $M$.

\section{Clustering Maximizing Information}

\subsection{Agglomerative Clustering}

An agglomerative clustering algorithm iteratively aggregates items from $X$ into groups, leading to a hard partitioning.
The algorithm starts with an initial partitioning where each item of $X$ is alone in its own cluster, i.e. $\mathcal{C}^{(X)} = \{c_i = \{x_i\}\}_{x_i \in X}$.
The clusters to merge are selected according to a cost function $D: C \times C \rightarrow \mathbb{R}^+$ which scores cluster pairs.
The pair $(c_i, c_j)$ with the lowest cost are merged together to form a new cluster $c_k = c_i \cup c_j$.
The agglomeration process take $n-1$ steps for the rows, where $n = |X|$.

The cost function affects the algorithm outcome \cite{clustering_agglo_fx, clustering_doc_retrieval}.
The selection of a specific cost function depends on the assumption over the cluster shape.
For instance, \textit{single linkage} focusses on merging clusters with the smallest gap $C(c_i, c_j) = \sum \min_{x_k \in c_i, x_{\ell} \in c_j} d(x_k, x_{\ell})$, without taking into account the cluster mass,  while \textit{complete linkage} focusses on merging clusters with the lowest maximal distance $C(c_i, c_j) = \sum \max_{x_k \in c_i, x_{\ell} \in c_j} d(x_k, x_{\ell})$.
It results in two different behaviors: \textit{single linkage} is sensitive to noise, as it would create artificial bridges between clusters, while \textit{complete linkage} is sensitive to outliers, preventing the merge of clusters containing some of them.

On the vector space model, the use of distance measures is not satisfactory \cite{feature_red_short_text}, as the information per documents is too short to get an accurate representation.
Instead of distance, the dissimilarity between clusters is measured using the divergence between their probability distribution.

A partitioning with one large cluster and many singleton clusters made of outliers
is similar to a filtering algorithm.
A clustering with such an outcome is not desirable as no \textit{true} group exists.
A partition must be composed of clusters with equivalent size, without high disparity.
Some algorithms naturally take into account the cluster size.
For example, in the case of \textit{complete linkage}, where the larger a cluster becomes, the harder it is to merge as the maximal distance to other clusters tends to grow.
When using a cluster probability distribution, all items within the clusters are represented by a single prototype independent of the cluster size.

To remediate to the fact that cluster prototypes do not include the knowledge of their size,
we define our agglomerative algorithm cost as the product between the cluster prototype divergence and the cost relative to their size:
\begin{equation} \label{eq:decision_cost}
    D^*(c_i, c_j) = D(c_i, c_j) \times {\rm Merge}(c_i, c_j; \mathcal{C})
\end{equation}
where $D(.)$ is the divergence part, while ${\rm Merge}(.)$ corresponds to the size part; this ensures that quality and quantity are similar across clusters.
These two parts will be defined in the following.

For simplicity, the \textit{features} are relatively defined  to the \textit{samples} considered.
When looking at rows, the features represented by columns, while when looking at columns, the relative features correspond to the rows.

\subsection{Partitioning Entropy}

\subsubsection{Shannon Entropy}

The Shannon entropy is a way to measure the number of bits required on average to code an information.
The more bits are needed, the more information would transit.

For a random variable with discrete values $X=\{x_i\}$ and associated probabilities $Pr(X=x_i) = p_i$, the Shannon entropy is defined as:
\begin{equation} \label{marginal_entropy}
  H(X) = - \sum_{x_i \in X} p_i \log p_i
\end{equation}
with the $\log$ corresponding to the base $2$ logarithm $\log_2$.

\subsubsection{Informative Clustering}

We state that a partitioning $\mathcal{C}$ is informative if items are distributed into clusters of equivalent size, maximizing the entropy $H(\mathcal{C})$.
Here, the probabilities associated to each cluster is not related to the item frequency, ie $Pr(C=c_i) \neq \sum_{x \in c_i} Pr(X=x)$.
Otherwise, the goal would be to isolate highly frequent keywords into individual clusters and gather infrequent keywords on a large cluster, which is by no means more interesting than clustering infrequent keywords alone.
Therefore, the cluster contribution is proportional to the number of items $|c_i|$,
$Pr(C=c_i) = p_i = \frac{|c_i|}{\sum_{c_j \in \mathcal{C}} |c_j|}$.
To make the distinction with entropy using item distribution, we refer to the entropy using item count as the \textit{partition entropy}.

For a partitioning into $k$ clusters, $H(\mathcal{C})$ is maximal for $Pr(C=c_i) = \frac{1}{k} \forall i$ with a maximal value of $\log k$.
The partitioning with the largest entropy is the one with a single item in each cluster.
To compare fairly two partitions at different stages of the agglomerative process regardless the number of clusters, the partition entropy is normalized by its theoretical maximum:
\begin{equation} \label{eq:relative_entropy}
  H_{rel}(\mathcal{C}) = \frac{H(\mathcal{C})}{\log|\mathcal{C}|}
\end{equation}
which is defined for any partitioning with at least $2$ clusters. The relative entropy takes values in $[0, 1]$, which allows convenient state comparison regardless the number of partitions.

\subsubsection{Partitioning Entropy Variation}

At the start, each item forms its own cluster, leading to a $H_{rel}(\mathcal{C}) = 1$.
The agglomerative process leads to clusters of various sizes which affects the entropy's quality.
To keep this value maximal, we study the entropy variation following a merge.
For two clusters $c_i$ and $c_j$  merged together, with probability $p_i$ and $p_j$, the new entropy can be expressed using the previous term:
\begin{equation} \label{eq:relative_entropy_new}
    \begin{array}{ll}
        H_{rel}(\mathcal{C}^{k-1})  &= H_{rel}(\mathcal{C}^k) \frac{\log k}{\log (k-1)} \\
         & + \frac{p_i\log p_i + p_j\log p_j}{\log k} - \frac{(p_i + p_j)\log (p_i + p_j)}{\log (k-1)} \\
         & = H_{rel}(\mathcal{C}^k) \frac{\log k}{\log (k-1)} + \Delta(p_i, p_j; k)
    \end{array}
\end{equation}
where $\mathcal{C}^{k-1}$ corresponds to the new partition with $k-1$ clusters and $\mathcal{C}^{k}$ the previous partition with $c_i$ and $c_j$ unmerged.
The total entropy is improved by a factor $\frac{\log k}{\log (k-1)}$ regardless of which clusters are merged.
Concerning the merged clusters contribution $\Delta(p_i, p_j; k)$, the behavior can be estimated for large $k$, as the approximation $\log k \approx \log (k-1)$ holds.
The merge impact is equivalent to $\Delta(p_i, p_j; k) \approx \frac{1}{\log k}\left(f(p_i) + f(p_j) - f(p_i+p_j)\right)$ where $f(x) = x \log x$.
$f$ is negative, concave and monotonically decreasing over the interval $[0, e^{-1}]$, with $e^{-1}=\exp(-1)$ corresponding to the limit of what could be considered as \textit{small} clusters, which is respected for many partitions as $3 e^{-1} > 1$.
As a consequence, $f(p_i) + f(p_j) < f(p_i + p_j)$ which leads to an entropy decrease, compensated to some extend by $H_{rel}(\mathcal{C}^k) \frac{\log k}{\log (k-1)}$.

\subsubsection{Cluster Size Influence}

For large $k$, $\Delta(p_i, p_j) < 0$.
To study the size influence, two merges are compared: the merge of $c_i$ and $c_j$ with respective probabilities $p_i$ and $_j$, and the merge of $c_i'$ with $c_j'$ with respective probabilities $\alpha p_i$ and $\alpha p_j$, where $\alpha \in \mathbb{R}^+$.
It can be shown that $\Delta(p_i', p_j') = \alpha \Delta(p_i, p_j)$.
The cost increases with the size of clusters merged.
To maximize the relative partitioning entropy cost over the agglomerative process, small clusters must be preferentially merged to limit the loss, which can be compensated by the former term  $H_{rel}(\mathcal{C}^k) \frac{\log k}{\log (k-1)}$ in eq. \ref{eq:relative_entropy_new}.

\subsubsection{Minimization Criterion}

For any pair of clusters $(c_i, c_j)$, the term $H^{rel}(\mathcal{C}) \frac{\log k}{\log (k-1)}$ in \ref{eq:relative_entropy_new} is the same regardless of which clusters are merged.
Two different merges are distinguished by the value of $\Delta(p_i, p_j; k)$.
As this term is negative, the goal is to minimize:
\begin{equation} \label{eq:end_merge_cost}
  {\rm Merge}(c_i, c_j; \mathcal{C}) = -\Delta(p_i, p_j; k)
\end{equation}

\subsection{Content Similarity}

We discussed about cluster size in the previous paragraphs.
The following describes the evaluation of clusters' content similarity.

\subsubsection{Cluster Conditional Probability}


Given a cluster $c^{(X)} \in \mathcal{C}^{(X)}$ and a partitioning $\mathcal{C}^{(Y)}$, the distribution of cluster $c^{(X)}$  over $\mathcal{C}^{(Y)}$ is:
\begin{equation} \label{eq:cluster_proba_distribution_X}
  Pr(c^{(Y)} | \mathcal{C}^{(X)}=c^{(X)}) = \frac{\sum_{x_i \in c^{(X)}} \sum_{y_j \in c^{(Y)}} M^*_{i, j}}{\sum_{x_i \in c^{(X)}}  \sum_{y_j \in Y} M^*_{i, j}}
\end{equation}
and for the distribution of cluster $c^{(Y)}$ over $\mathcal{C}^{(X)}$:
\begin{equation} \label{eq:cluster_proba_distribution_Y}
  Pr(c^{(X)} | \mathcal{C}^{(Y)}=c^{(Y)}) = \frac{\sum_{x_i \in c^{(X)}} \sum_{y_j \in c^{(Y)}} M^*_{i, j}}{\sum_{x_i \in X}  \sum_{y_j \in c^{(Y)}} M^*_{i, j}}
\end{equation}

\subsubsection{Cluster Dissimilarity}

For two clusters $c_a, c_b \in \mathcal{C}^{(X)}$, the probability distribution over $\mathcal{C}^{(Y)}$ is noted $A$ and $B$ respectively to limit notation symbols, such as $a_i = Pr(c^{(Y)}_i | c^{(X)}_a)$ and $b_i = Pr(c^{(Y)}_i | c^{(X)}_b)$.
The Kullback-Leibler ($KL$) divergence is a way to measure the distance between probability distributions $A$ and $B$:
\begin{equation} \label{KL}
    KL(A \| B) = \sum_i a_i \log\frac{a_i}{b_i}
\end{equation}
The same equation is obtained for $c_a, c_b \in \mathcal{C}^{(Y)}$, with $a_i = Pr(c^{(X)}_i | c^{(Y)}_a)$ in this case.
The intuition of the $KL$ divergence is that mass of distribution $B$ must be present where $A$ is.
If not, the penalty grows.
The $KL$ divergence can be rewritten as $KL(A \| B) =  H^*(A, B) - H(A)$ where $H^*(A, B)$ is the cross-entropy, and $H(A)$ the regular entropy.
While the entropy $H(A)$ corresponds to the average number of bits exchanged to communicate symbols of $A$ using the most optimal code, the cross-entropy $H^*(A, B)$ corresponds to the average number of bits exchanged to transmit $A$ given the optimal code to transmit $B$.
If the two clusters share the same distribution, the code is likely to be similar, and the associated cost low.

\subsubsection{Symmetry}

There is no order when merging two clusters, as $c_i \cup c_j = c_j \cup c_i$.
However, the Kullback-Leibler divergence is not symmetric.
Instead, we use the $J$-symmetrized $KL$ divergence, which is defined as:
\begin{equation} \label{J_KL}
    KL^J_{\alpha}(A \| B) = (1 - \alpha)KL(A \| B) + \alpha KL(B \| A)
\end{equation}
with the balance factor $\alpha = \frac{1}{2}$.
The symmetrization ensures that both $A$ and $B$ share the same support probability, which leads to a more discriminative function as both $a_i$ and $b_i$ needs to be non-zero for the same feature $i$.

\subsubsection{Minimization Criterion}

The cost term takes into account the clusters similarity in eq. \ref{eq:decision_cost} is
$D(c_a, c_b) = KL^J(A \| B)$.
It measures the divergence between prototypes' distribution according to the clustered features.
The $KL^J$ takes values in $\mathbb{R}^+$, where a low value represents a high content similarity between considered clusters.

\subsection{Agglomeration Procedure}

Given the initial smoothed documents-keywords matrix $M^*$,
the algorithm starts by computing $KL^J$ for all possible pairs of clusters in  $\mathcal{C}^{(X)}$ and pairs in  $\mathcal{C}^{(Y)}$.
This lead to two initial divergence matrices $KL^J(\mathcal{C}^{(X)})$ and $KL^J(\mathcal{C}^{(Y)})$.
This operation is computationally expensive, as it requires $\mathcal{O}(nm (m + n))$ operations.

The merge cost is recomputed at each round, and the total cost is computed for each pair.
The pair of clusters from $\mathcal{C}^{(X)}$ or $\mathcal{C}^{(Y)}$ with the lowest cost is selected and merged.
The two divergence matrices are updated after the merge operation.
If two clusters $c_i, c_j \in \mathcal{C}^{(X)}$ are merged together, all the pairs involving $c_i$ and $c_j$ in $KL^J(\mathcal{C}^{(X)})$ must be recomputed with the new cluster characteristics $c_i \cup c_j$, leading to a matrix with one dimension less.
This first update requires $\mathcal{O}(n(t)m(t))$ operations,  where $n(t)=|\mathcal{C}^{(X)}|$ and $m(t)=|\mathcal{C}^{(Y)}|$ is the number of row and column clusters left after $t$ merge operations.

Concerning $KL^J(\mathcal{C}^{(Y)})$, all the items are affected.
Nonetheless, the matrix can easily be updated, by looking at the difference between merged and unmerged state.
For two column clusters $c_a$ and $c_b \in \mathcal{C}^{(Y)}$ with distribution over rows $A$ and $B$ respectively, the cost variation is:
\begin{equation}\label{eq:delta_reciprocal}
  \begin{array}{llc}
      \Delta^{(i, j)}KL(A \| B)  & =
       (a_i + a_j) \log \frac{a_i + a_j}{b_i + b_j} \\
      & - \left( a_i \log \frac{a_i}{b_i} + a_j \log \frac{a_j}{b_j} \right)  \\
  \end{array}
\end{equation}
where $\Delta^{(i,j)}KL(A\|B)$ corresponds to the non-symmetric $KL$ cost.
The new cost for merging $c_a$ with $c_b$ is replaced by $KL^J(c_a, c_b) + \Delta^{(i, j)}KL(A \| B) + \Delta^{(i, j)}KL(A \| B)^{\dagger}$.
This updating step requires $m(t)^2$ operations to update all the feature pairs.
In total, a step requires $\mathcal{O}(n(t)^2 + m(t)^2)$ operations to select the best pair, and $\mathcal{O}(n(t)m(t) + m(t)^2)$ operations to update the cost pairs $KL^J$  when merging two row clusters.
The same reasoning applies when merging two-column clusters by exchanging $n(t)$ with $m(t)$ in the formula.

\section{Experimental Setup}

\subsection{Datasets}

\subsubsection{Scientific Paper Tags}

The main motivation for tag co-clustering arose from scientific papers literature.
The DBLP dataset \cite{Tang:08KDD} is a citation graph gathering computer science papers, with meta-data such as title, publication year, references, authors, conference/journal, \textit{field of study}, available for a large number of papers.
We used the most recent version (v12) for our experiments.

The \textit{field of study} is a list of descriptive keywords about the field (e.g. \textit{Cryptography}, \textit{Biology}), the method (\textit{Matrix factorization}), or other related concepts (\textit{Bullwhip  effect}) discussed in a paper. Each paper has, on average $10$ descriptive tags.
Hopefully, tags are already well pre-processed, and no steaming nor stop-words removal need to be done.

The algorithm complexity is more than quadratic, which prevents the scaling to a large database.
Documents are sampled to make the clustering possible on a regular machine.
A particular tag is selected, and all papers with the tag included are gathered.
Then, $5000$ documents are selected at random from this pre-selection.
All keywords with less than $5$ occurrences are discarded, which leads to around $1000$ keywords left and a filling rate of $2 \%$ of the binary documents-keywords matrix.

\subsubsection{Synthetic checkerboard}

The real-world dataset does not contain any label, which prevents the evaluation with objective metrics.
We propose to generate a sparse synthetic dataset with clustering structures to test the performance of our model.

A synthetic sparse matrix $M$ of size $n_X \times n_Y$ partitioned over $X$ and $Y$ dimensions is constructed the following way.
Rows are split into $k_X$ clusters of equal size $\lfloor \frac{n_X}{k_X} \rfloor + \{0, 1\}$.
The same regular partitioning is performed on $Y$ with $k_Y$ clusters.

The matrix $M$ is filled with $0$ and $1$ according to the partitioning.
The  \textit{tile} $T(a, b) = \{M^*_{i, j}\}_{ x_i \in c_a^{(X)}, y_j \in c_b^{(Y)}}$ is the intersection between the row and column clusters  $c_a^{(X)}$ and $c_b^{(Y)}$.

Some of the tiles selected with probability $\alpha \in [0, 1]$ are filled, leaving $(1 - \alpha)$ of the tiles empty, where $\alpha$ is the \textit{global} filling rate.
For a tile $T(a, b)$ to fill, the filling rate $\beta_{a, b}$ is selected at random in $[0, \beta]$ with $\beta \in [0, 1]$ the \textit{local} filling rate.
For each item $(i, j)$ in tile $(a, b)$ to be filled, its value is $1$ with probability $\beta_{a, b}$ else $0$.
The result is a matrix filled with rate $\frac{\alpha \beta}{2}$.

In our case, the global and local filling rate are set to $\alpha = \beta = 0.2$ leading to a total filling rate similar to our real-world dataset of $2 \%$ .
For all experiments, $n_X = n_Y = 1000$ and $k_X = k_Y$ would be adjusted over the experiments.
The resulting matrix looks like a regular grid and would be called, for this reason, the \textit{checkerboard} dataset (see Fig. \ref{fig:smoothing}).
The experiments are done for the same $n_X$ and $n_Y$, and identical $k_X$ and $k_Y$.
This choice is made to aggregate results over $X$ and $Y$ together, but the performances are not affected by asymmetric choices.

\subsection{Monitoring metrics}

To evaluate our algorithm, we selected some supervised and unsupervised metrics to evaluate the quality of the partition recovery and estimate cluster quality in the absence of labels.

\subsubsection{V-measure}

The $V$-measure is a supervised metric comparing the real clusters to the estimated ones using entropy measures.
It is analogous to \textit{accuracy} on classification problems.
Two sub-measures are first computed: the \textit{homogeneity}, which corresponds to the fact that a good cluster contains a single class, and the \textit{completeness}, which measures how well elements from a given class are grouped.

The homogeneity is defined as:
\begin{equation}\label{homogeneity}
  h = 1 - \frac{H(L | K)}{H(L)}
\end{equation}
with $L$ the real cluster labels and $K$ the hypothetic labels obtained using a clustering.
The completeness is defined similarly as:
\begin{equation}\label{completeness}
  c = 1 - \frac{H(K | L)}{H(K)}
\end{equation}
The $V$-measure is then defined as:
\begin{equation}\label{v_measure}
  V = \frac{(1 + \beta)\times h  c}{\beta h + c}
\end{equation}
where the parameter $\beta \in \mathbb{R}^+$ balances the contribution of each term.
For $\beta=0$, it corresponds to the $h$, while $\lim_{\beta \rightarrow \infty} = c$. The values obtained lie within $[0, 1]$, where $1$ is attributed to the best clustering, while $0$ to the worse case.

The $V$-measure can also be used to compare two partitioning obtained with different parameters or algorithms, measuring the similarity degree.

\paragraph{Random Guess}

Suppose $\mathcal{C}$ is a partitioning of items into $k$ clusters of equal size, with $Pr(c) = \frac{1}{k}$.
The associated partitioning entropy is $H(\mathcal{C})= \log k$.
Be $\mathcal{C}'$ a randomly guessed partitioning with $k$ clusters of equal size too, but filled with items selected at random.
The overlap probability between $c \in \mathcal{C}$ and $c' \in \mathcal{C}'$ is $\frac{1}{k^2}$ for all clusters' pairs.
Consequently, the joint entropy is $H(\mathcal{C}, \mathcal{C}') = \log k^2 = 2 \log k$, leading to $H(\mathcal{C} | \mathcal{C}') = H(\mathcal{C}' | \mathcal{C}) = \log k$.
Completeness and homogeneity are both equal to 0, leading to an undefined $V$-measure.
However, when looking at the limit, the value converges to zero.
Compared to accuracy measure, where a random guess's accuracy is  $\frac{1}{k}$, the $V$-measure is more discriminative.

\subsubsection{Limited Partitioning Entropy}

At the start, the partitioning entropy defined in eq. \ref{eq:relative_entropy} is maximal.
However, the clustering is non-informative as only singleton clusters exist.
Instead, knowing that final clusters would have a critical size with more than $r$ elements, smaller clusters' contribution  can be discarded, considering them as \textit{outliers}.
The restricted relative partitioning entropy is then defined as:
\begin{equation} \label{eq:relative_entropy_restricted}
  H^*_{rel}(\mathcal{C}; r) = \frac{- \sum_{c \in \mathcal{C} \wedge |c| > r} p(c)\log p(c)}{\log |\mathcal{C}|}
\end{equation}
At initialization, the value is $0$ as no cluster of sufficient size exists for $r > 1$.
This value is still bounded between $[0, 1]$ and enables to track clusters creation. This measure allows to evaluate the partitioning distribution without considering cluster content.

\subsubsection{Mutual Information}

When  monitoring the cluster's content, the information variations are monitored.
In this case, the entropy is computed using the sample probabilities,
defined in equations \ref{eq:cluster_proba_distribution_X} and \ref{eq:cluster_proba_distribution_Y}.
The \textit{mutual information} corresponds to the information shared between $X$ and $Y$.
This measure is defined as:
\begin{equation} \label{mutual_info}
    I(X, Y) = H(X) + H(Y) - H(X, Y)
\end{equation}
It corresponds to the gain of coding $X$ with $Y$, i.e. the amount of redundancy between the two variables.
This value is bounded by $0$ (independent variables) and $\min(H(X), H(Y))$ (correlated variables).
We would monitor the mutual information between partitioning $I(\mathcal{C}^{(X)},\mathcal{C}^{(Y)})$.
As the number of clusters decrease over time, this value would decrease due to information loss during the compression.

\subsection{Comparative Algorithms}

We proposed to compare our algorithm to the co-clustering algorithm presented in \cite{Dhillon01co-clusteringdocuments}, which relies on spectral decomposition.
First, two diagonal matrices $D_{1}$ and $D_2$ are obtained from $M$, where $D_{1: i,i} = \sum_{j=1}^m M_{i, j}$ and $D_{2: j,j} = \sum_{i=1}^n M_{i, j}$.
Then, the normalized matrix  $M_n = D_1^{-1/2} M  D_2^{-1/2}$ is decomposed using singular value decomposition such as $M_n = U S V^T$.
The vectors $D_1^{-1/2}U$ and $D_2^{-1/2} V$  are concatenated to form the matrix $Z$.
The algorithm finishes by performing a $k$-means clustering  on the $\ell = \lceil \log_2 k\rceil$ main dimensions, omitting the first main dimension.

\pagebreak

\section{Experimental Results}

\subsection{Smoothing Effect}

\begin{figure}
  \centering
  \includegraphics[width=\columnwidth]{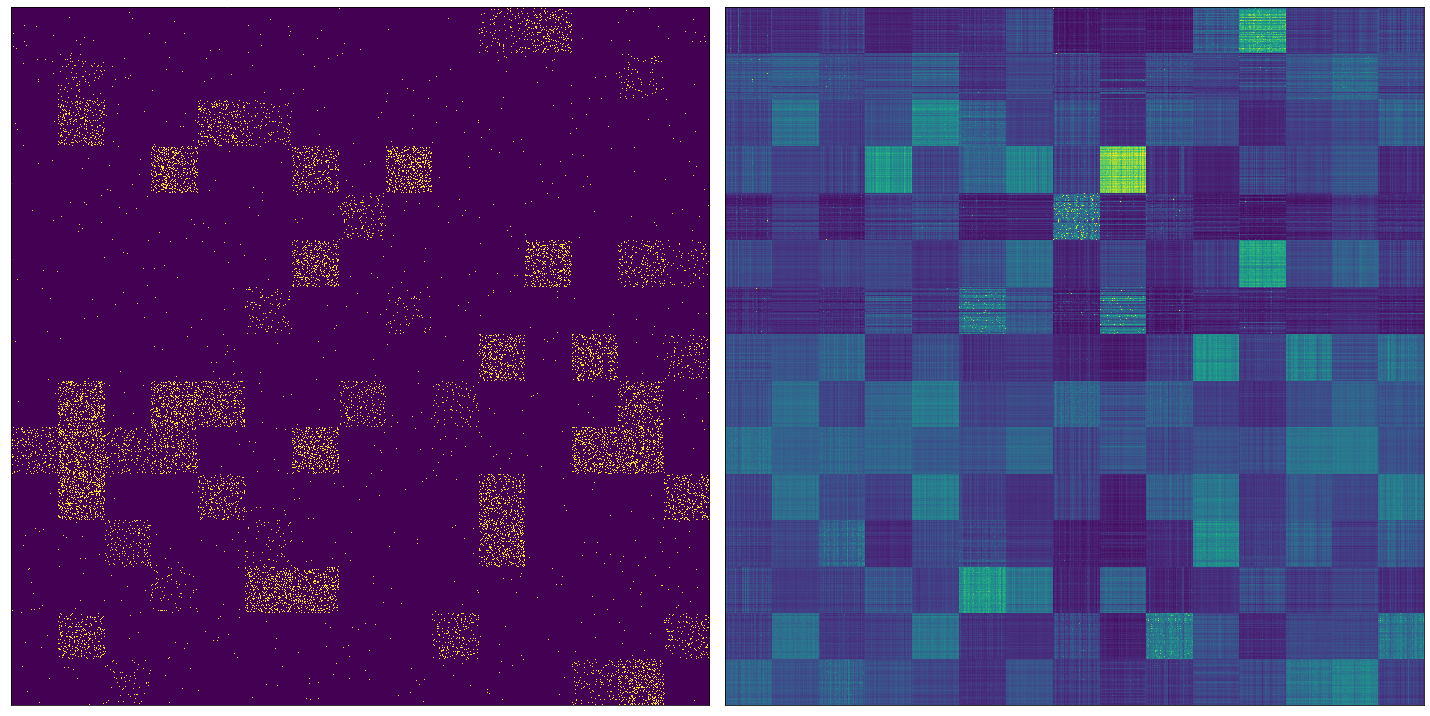}

  \caption{Matrix smoothing for $n_X=n_Y=1000$ and $k_X=k_Y=15$, ie around $67$ points per cluster. The global filling ratio is $\alpha=0.2$ and the local filling ratio is $\beta=0.2$. Left: binary matrix, right: smoothed matrix.}
\label{fig:smoothing}
\end{figure}
The smoothing proposed enables to switch from binary values to real values by redistributing the weights.
A visual example is presented in Fig \ref{fig:smoothing}.
The filled tiles are identifiable on the binary matrix.
For the tiles with low $\beta_{a, b}$, the boundaries are hard to identify.
On the smoothed matrix, the weights are completely redistributed leading to the visual identification of tiles, even the ones that were not filled at all.
All items belonging to the same clusters tend to have a more similar feature vector.
The information is of lower intensity locally but is better distributed across the different features, even on tiles that were not filled.

\subsection{Size Dependent Cost}

This experiment compares the composite cost defined in eq. \ref{eq:decision_cost} to the version where only content similarity obtained using $KL^J$ cost is taken into account.
For this purpose, we compared for several numbers of cluster $k$ the maximal $V$-measure averaged over $5$ independent trials for each $k$.
We gathered the number of clusters left $\hat{k}$ for the maximal value of $V$.

We did the same by extracting the maximal restricted relative entropy, removing clusters of size smaller or equal to $1$,  and extracting the corresponding $\hat{k}$.
\begin{figure}
  \centering
  \includegraphics[width=\columnwidth]{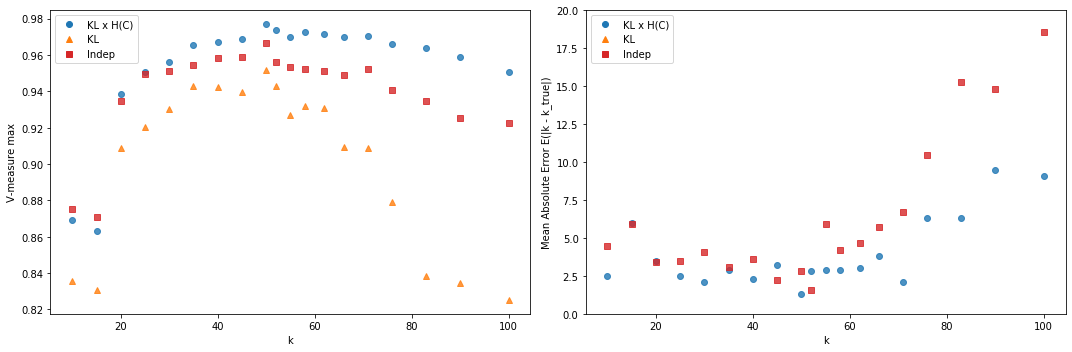}

  \caption{Best $v$-measure and average absolute error for corresponding $k$.
Blue rounds correspond to the normal setup with co-clustering and weighting according to $KL^J$ cost and cluster distribution entropy $H(\mathcal{C})$.
Orange triangles correspond to the co-clustering setup without taking into account cluster distribution.
Red squares correspond to the double weighting approach but rows and columns are clustered independently.
Each point corresponds to the average for $5$ independent trials.
}
\label{fig:v_measure_setup_1}
\end{figure}
The results are presented in Fig. \ref{fig:v_measure_setup_1}, with the composite cost denoted $KL \times H(\mathcal{C})$ and the simple cost $KL$.

As a general remark, for all setups, it is easier to cluster many clusters with sufficient size.
For very few clusters, the accuracy quickly decreases.
As the global filling rate is $\alpha=0.2$, on average $2$ tiles are filled over $10$.
Due to randomness, a single one or none could be filled, leading to less information for clustering.
For a large number of partition, a cluster is well defined in the sense that enough tile are filled.
The problem of insufficient information occurs at the sample level.
For a row sample of size $n$ and $k$ feature partitioning, there are $\frac{n}{k}$ slot for a given feature cluster.
As the local filling rate is $\beta_{a, b} \leq 0.2$, the probability that none of the slot are filled grows with $k$. This effect is nonetheless less disturbing than the former as redundancy exists.

The $V$-measure of the composite cost is always higher than for the simple cost setup.
The accuracy of $KL$ quickly drops for large $k$  with smaller cluster sizes.
The absolute $k$ deviation $\mathbb{E}(|k - \hat{k}|)$ of the composite cost is relatively close from the optimal, with a consistent error unless for very small cluster size, while the error for the simple cost is too large to fit in the figure.
\begin{figure}
  \centering
  \includegraphics[width=\columnwidth]{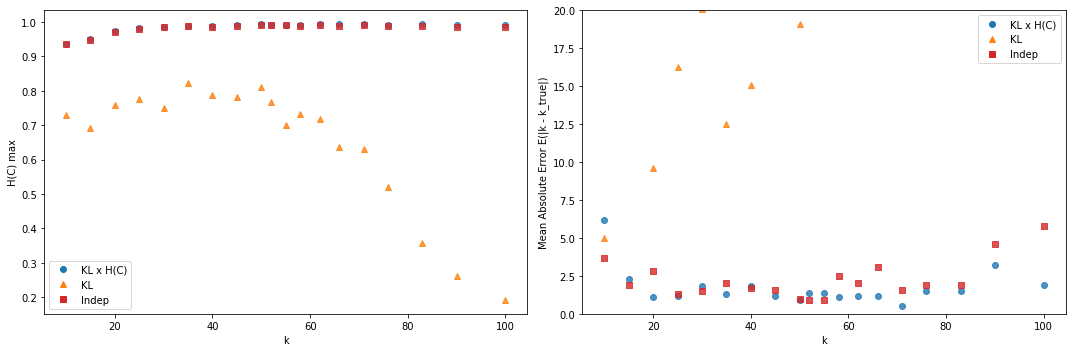}
  \caption{Left: Best $H(\mathcal{C})$ with cluster of size lower or equal to $1$ removed.
  Right:  average absolute error of the corresponding $\hat{k}$.
  Same color code and setup as \ref{fig:v_measure_setup_1}.
  Blue rounds and red circles overlap on the left sub-figure.
}
\label{fig:entropy_setup_1}
\end{figure}
Looking at the maximal relative partitioning entropy in Fig. \ref{fig:entropy_setup_1}, the composite cost leads to very good cluster distributions for any $k$, while the simple cost is $0.2$ points lower for small number of clusters and very low for larger values.
On the right side of Fig. \ref{fig:entropy_setup_1}, the $\hat{k}$ obtained with the composite cost is close to the exact value even for large $k$.
This means that the number of clusters obtained when stopping the agglomeration procedure with the partitioning entropy criterion is close from the true initial cluster numbers for all cluster sizes.
When using the simple cost, clusters number is far from the true number of cluster.
In both case, the entropy stopping criterion leads to a smaller error over the number of estimated clusters.

\subsection{Co-Clustering vs Independent Clustering}

One of the initial hypothesis concerns the synergy between joint reduction.
We compare the co-clustering setup to the independent setup, where the partitioning $\mathcal{C}^{(X)}$ is obtained using the uncompressed features $Y$, as well as the partitioning  $\mathcal{C}^{(Y)}$ is obtained using the unaggregated rows $X$.
This setup is denoted \textit{Indep} in Fig. \ref{fig:v_measure_setup_1} and \ref{fig:entropy_setup_1}.

The results obtained with the independent clustering are similar to the co-clustering but with a lower $v$-measure for a large number of clusters $k$.
The obtained $\hat{k}$ from the $v$-measure are close to the co-clustering ones.
As far as shape is concerned, independent clustering performs as well as the co-clustering, and the $\hat{k}$ obtained are relatively similar.
The  co-clustering advantage is limited for large clusters / small $k$ and becomes more interesting when uncertainty grows with smaller size clusters for large $k$.

\subsection{Comparison to Alternative Algorithms}

\begin{figure}
  \centering
  \includegraphics[width=\columnwidth]{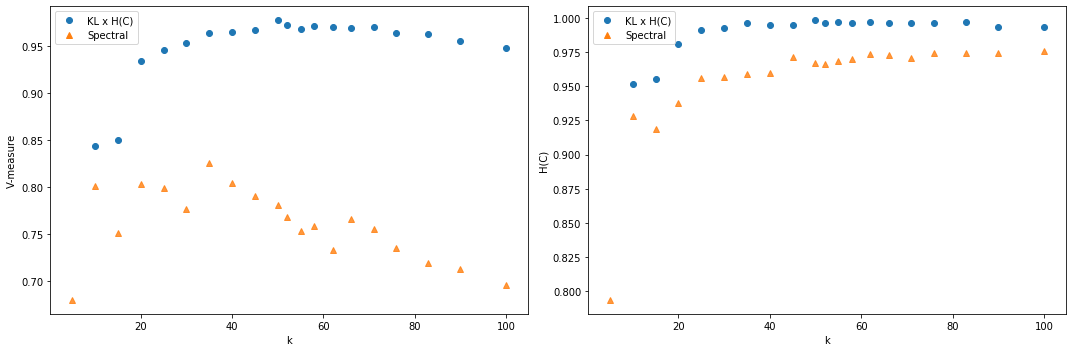}

  \caption{Comparative results between Spectral co-clustering and Agglomerative clustering. On the right, maximal $V$-measure obtained for
}
\label{fig:spectral_vs_hierarchical}
\end{figure}

We compare our algorithm to spectral co-clustering presented in \cite{Dhillon01co-clusteringdocuments}.
The algorithm needs as input the target number of clusters to search for.
When comparing the agglomerative algorithm with the spectral algorithm,
the spectral algorithm is run with the exact $k$ provided,
compared to the partitioning obtained with the agglomerative algorithm with $k$ remaining clusters.
The $V$-measure and the relative cluster partitioning entropy $H_{rel}^*(\mathcal{C}; 1)$ are extracted from these two partitioning.
The results are presented in Fig. \ref{fig:spectral_vs_hierarchical}.
The spectral algorithm results are lower than the one obtained for the agglomerative approach.
However, the shape of the clusters obtained are equivalent.
The spectral approach is quite robust in general, but the sparsity level affects the results.
With a higher filling rate ($\alpha=0.4$), the spectral results get closer to the agglomerative one.

We also tested with DBSCAN, which has been used in some papers.
As it is impossible to select the wished number of clusters, and because the results were lower than spectral decomposition, the results are not presented.
Nonetheless, the smoothed matrix's use improved the partitioning, allowing the algorithm to discover more clusters than with the regular binary matrix.

\subsection{Textual Results}

The initial goal was to cluster tags associated with scientific papers to identify topics.
In the dataset used, there is no high-level classification or paper grouping to evaluate our clustering.
Despite the lack of objectivity, we present the results on two subsets of papers, obtained for the \textit{Payment} field of study, and the second for \textit{Biometry}.

We take advantage of the hierarchical form to present the results using a dendrogram.
Around $15$ clusters are left unmerged, and the three most frequent keywords are displayed for analysis.
For the two, the relative partitioning entropy was around $H(\mathcal{C})=0.95$ for keywords.

\paragraph{Payment}

\begin{figure}
  \centering
  \includegraphics[width=\columnwidth]{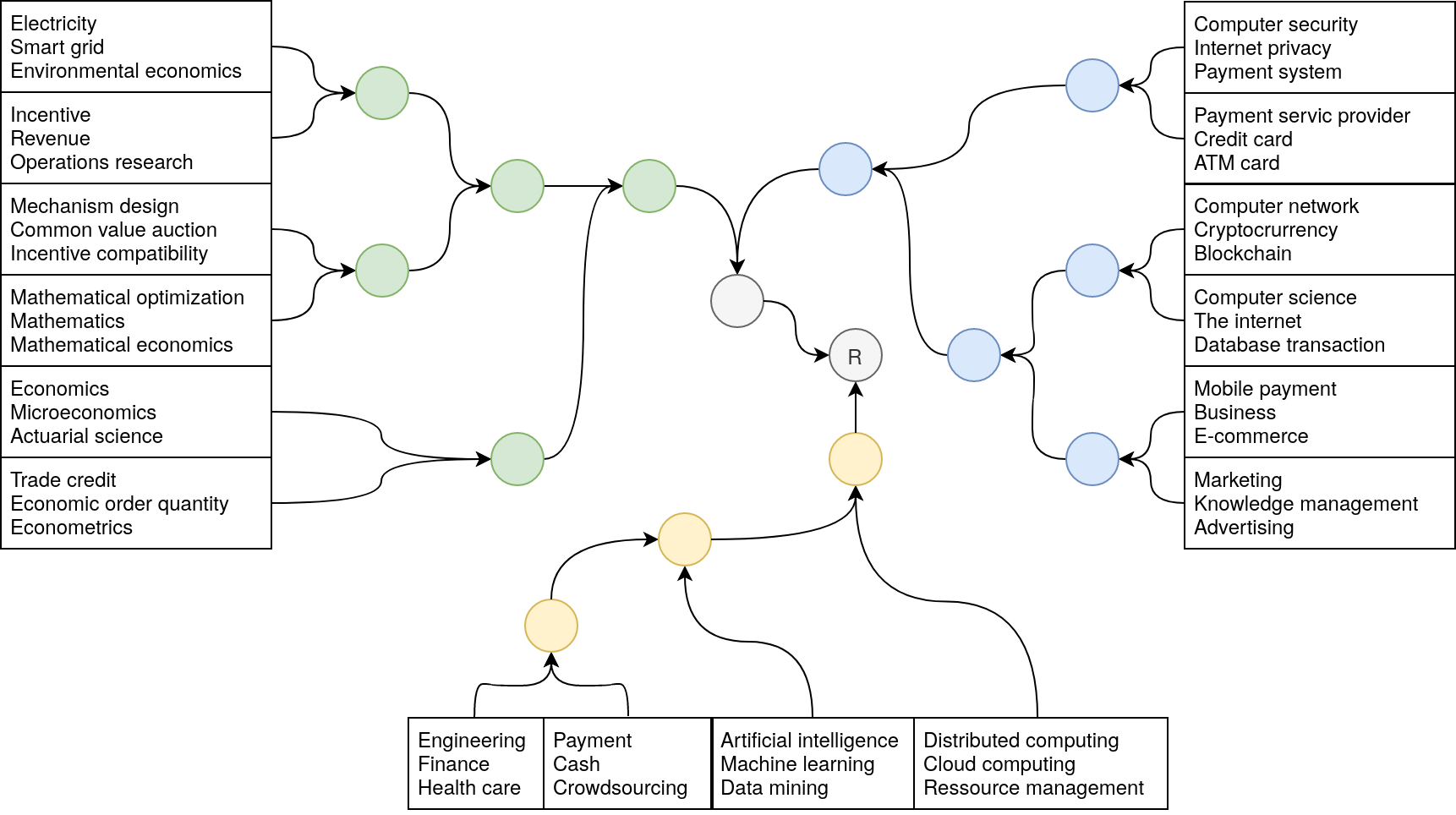}
  \caption{Dendrogram for \textit{Payment} field of study.}
\label{fig:dendrogram_payment}
\end{figure}
Fig. \ref{fig:dendrogram_payment} corresponds to the clustering of keywords co-occurring with the \textit{Payment} tag.
Three high level clusters are identified.
The one on the left corresponds to things related to \textit{economics}.
The right one corresponds to what could be considered as the \textit{core} of the payment field, oriented toward users, with the new payment methods (\textit{Cryptocurrency},  \textit{Mobile payment}) and intricated topics (\textit{Computer security}, \textit{Marketing} and \textit{Advertising}).
The bottom cluster corresponds to the medium or technology used in the payment but is not specific.
For instance, \textit{Artificial intelligence} is used in payment systems for fraud detection or biometric authentication, but it is not specific to payment.

Surprisingly, the keyword \textit{Payment} is located on the bottom cluster, near \textit{Cash} and \textit{Crowdsourcing}, which seems conceptually incorrect.
This is due partially to our sampling method, where all documents with keywords  \textit{Payment} were selected.
As it co-occurs with all keywords, there is no way to identify true relationships. \textit{Payment} is located on a cluster were the other keywords are related to \textit{Crowds}, with additional keywords such as \textit{Reputation}, \textit{Social network}, \textit{Audit} and \textit{Crowdsensing}.

This artefact is not limited to the selected keywords but to the most frequent keywords.
A second example is \textit{Computer science} on the right, in a cluster related to the \textit{Internet}, with additional keywords like the \textit{World Wide Web}, \textit{Mobile device}, \textit{Service provider} and \textit{Mobile computing}.

\paragraph{Biometry}

\begin{figure}
  \centering
  \includegraphics[width=\columnwidth]{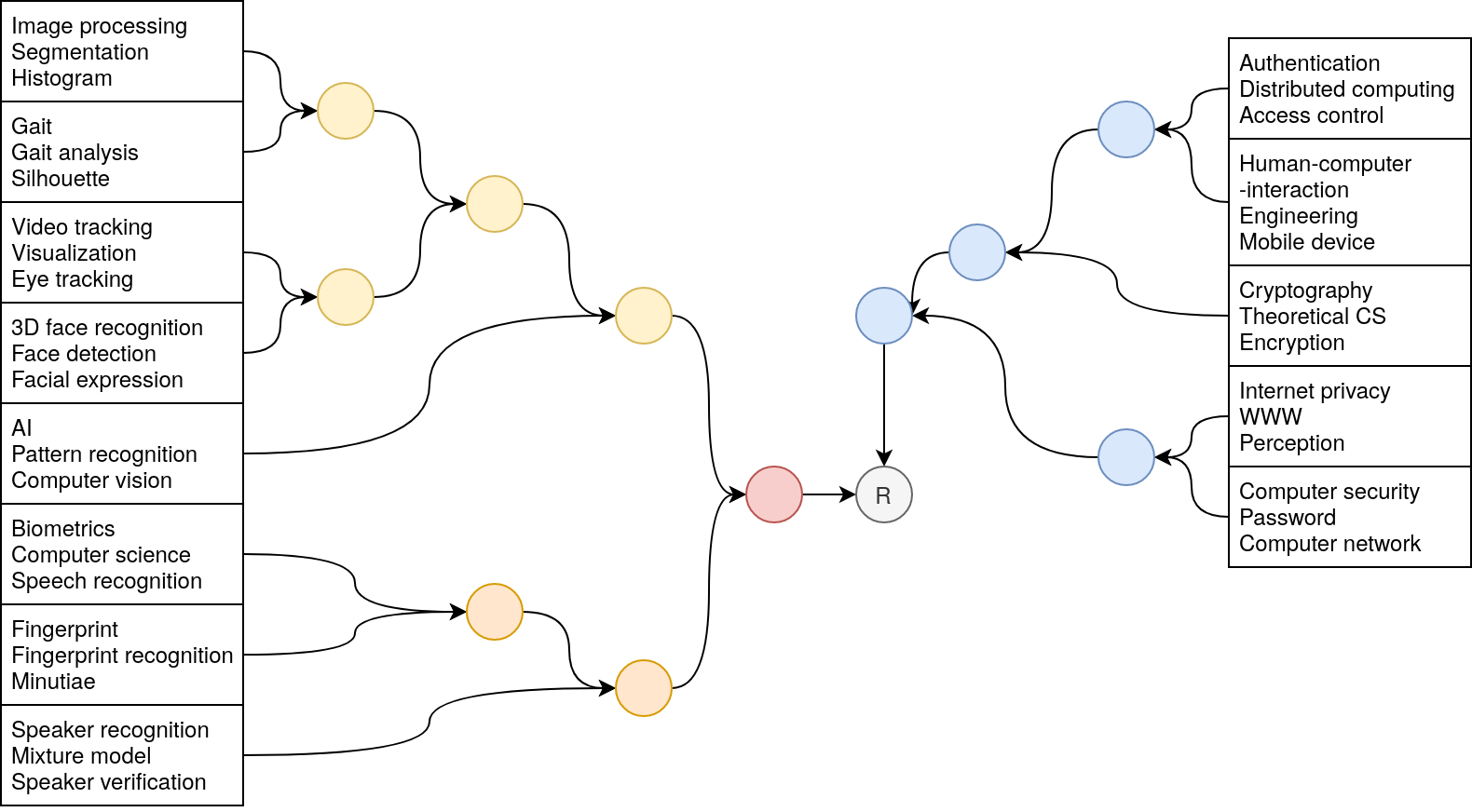}

  \caption{Dendrogram for \textit{Biometrics} field of study. }
\label{fig:dendrogram_biometry}
\end{figure}
The second partitioning uses the \textit{Biometrics} tag as a reference.
The resulting dendrogram is presented in Fig. \ref{fig:dendrogram_biometry}.
Two large clusters are identified.
The main on the left gathered keywords about biometric methods and algorithms to extract a digital identity.
It is subdivided into two subgroups.
The top one gathered keywords related to computer-vision, with \textit{Image processing}, gait and face analysis.
The bottom one corresponds to the other methods, with fingerprint, \textit{Speaker recognition}/\textit{verification}.
The cluster with \textit{Biometrics}, \textit{Computer science} and \textit{Speech recognition} corresponds to an artefact gathering highly frequent keywords together.
The right cluster corresponds to the security part, with \textit{Cryptography}, \textit{Password}, \textit{Authentication} and others.

\section{Discussions}

\subsection{Ending Criterion}

The checkerboard experiments were evaluated, knowing the number of clusters.
In an unsupervised setup, this knowledge is often unavailable.
To select the cluster number, one has to look at a specific criterion indicating if the partitioning is satisfying.
For instance, the algorithm $X$-means \cite{Pelleg00x-means} is a divisive algorithm based on $k$-means which successively splits the existing clusters.
The splitting decision is based on the split's likelihood, assuming the data corresponds to a Gaussian mixture.
For more general clustering algorithm, the silhouette coefficient, measuring the distance to the nearest cluster versus the radius of the cluster.

On our type of data, the silhouette is not suitable as cluster are not well separated.
The \textit{goodness} criterion of the algorithm must be in accordance to the goal achieved by the algorithm.
Reminding our cost definition in \ref{eq:decision_cost}, it is the product between \textit{cluster size} and \textit{content} related costs.

The first part of the answer to this problem is to look at the restricted relative partitioning entropy defined in \ref{eq:relative_entropy_restricted}, with small clusters of size $1$ or less removed.
In Fig. \ref{fig:entropy_setup_1}, $H_{rel}(\mathcal{C}; 2)$ is already a good indicator of when clusters are sufficiently aggregated.
However, this is a particular case where all clusters have the same size, leading to a particular configuration where $H_{rel}$ is maximal.
In a more general configuration, there is no particular reason for clusters of exactly the same size.

The second cost part takes into account content.
On the information theory-based work of \cite{Dhillon03InformationCoclustering},  a  good clustering is defined as minimizing the quantity $I(X, Y) - I(\mathcal{C}^{(X)}, \mathcal{C}^{(Y)})$ for a given number of row and column clusters.
The agglomeration of clusters is a form of compression which mechanically reduces the information available.

The restricted relative entropy is maximal towards the end of the agglomeration process, while the information is maximal at the beginning and minimal at the end.
A good compromise between the two is to look for the value for which the product of  $I(\mathcal{C}^{(X)}, \mathcal{C}^{(Y)})$ and $H(\mathcal{C}^{(X)})$ or $H(\mathcal{C}^{(Y)})$ is maximal:
\begin{equation} \label{eq:ending_criterion}
  k^*_X = \arg_{k_X = |\mathcal{C}|} \max H_{rel}(\mathcal{C}^{(X)}; r) \times I(\mathcal{C}^{X}, \mathcal{C}^{Y})
\end{equation}
$X$'s best partitioning is not necessarily simultaneous with that of $Y$ because the actual number of clusters may be different.
The co-clustering only exploit synergies to find clusters more accurately.
In general, the partitioning with the lowest number of dimensions would be merged more frequently until reaching a size comparable to its feature size.
As a rough guide, for $k$ features, the maximum entropy is $\log k$.
The cost of $KL^J$ is not limited by an upper bound, but the higher the number of  features, the higher the cost will be because the probability of $a_i=0 \neq b_i$ is higher in such a configuration.
With a higher cost, the clustering will preferably select the cluster pairs with the smallest number of features.

This criterion was tested on the checkerboard dataset.
The estimated $\hat{k}$ were very close to the expected one, with an average absolute deviation close to $1$.
This criterion was used to build the dendrogram, where the estimated cluster number was between $15 \sim 20$ clusters depending on the main selected keyword.

\subsection{Model Limitations}

The checkerboard model differs from a documents-keywords matrix obtained from tag on two majors points.

The first difference concerns distribution.
Tags follow a Zipf's law, which is not modeled here, as all columns have on average the same strength.
Nonetheless, the unbalanced distribution is corrected by the matrix smoothing protocol, which decreases the weight of these frequent keywords to less frequent one.

The second difference is the hierarchical division of keywords.
The associated tags range from very broad domains (\textit{Chemistry}, \textit{Mathematics}), to fields (\textit{Inorganic chemistry}, \textit{Databases}), to specialities and other lower levels.
The checkerboard model is made of independent clusters which are not hierarchicaly organized.
Nonetheless, due to the scaling limitation of the proposed algorithm, the restriction to around $5000$ documents and $1000$ keywords limit the visibility of such organizations.

\section{Conclusion}

In this paper, we addressed the problem of tag clustering, where the tag amount per document is limited.
To this purpose, using the correlation between tags and keywords, a method to enhance context without the use of an external database or model was proposed.
With the assumption that a clustering is informative if the partition entropy is large, we proposed an agglomerative co-clustering algorithm taking into account the content as well as the cluster size.
The algorithm showed good recovery performance on synthetic datasets with the same sparsity level.
It showed conceptually correct clustering results on scientific paper tags, up to highly frequent keywords where no discriminative relationship could be found.
The algorithm's complexity is polynomial but more than quadratic, which restricts its usage on a small dataset.
Some improvement can be made by splitting the dataset into independent parts or finding cost approximations.
Nonetheless, the idea of building groups of equivalent size could be mixed with other agglomerative measures to include  distant items to their closest cluster.

%
%

\bibliographystyle{unsrtnat}

\bibliography{references}  

\end{document}